\documentclass[referee]{aastex}

\usepackage{graphicx}
\usepackage{amssymb}
\usepackage{natbib}
\usepackage{amsmath}
\usepackage{color}

\newcommand{\myemail}{mat@igam.uni-graz.at, asv@igam.uni-graz.at}
\newcommand{\degree}{^{\circ}}

\shorttitle{CME acceleration and flare particle acceleration}
\shortauthors{Temmer et al.}

\begin{document}

\title{Combined STEREO/RHESSI study of CME acceleration and particle acceleration in solar flares}


\author{M. Temmer and A.M. Veronig}
\affil{IGAM/Kanzelh\"{o}he Observatory, Institute of Physics, Universit\"at Graz, Universit\"atsplatz 5, A-8010 Graz, Austria}\email{\myemail}

\author{Kontar E.~P.}
\affil{Department of Physics and Astronomy, University of Glasgow G12 8QQ, UK}\email{eduard@astro.gla.ac.uk}

\author{Krucker S.}
\affil{Space Sciences Laboratory, University of California, Berkeley, CA94720-7450}\email{krucker@ssl.berkeley.edu}

\author{B. Vr\v snak}
\affil{Hvar Observatory, Faculty of Geodesy, University of Zagreb,
Ka\v{c}i\'{c}eva 26, HR--10000 Zagreb, Croatia}
\email{bvrsnak@gmail.com}

\begin{abstract}
Using the potential of two unprecedented missions, STEREO and RHESSI, we study three well observed fast CMEs that occurred close to the limb together with their associated high energy flare emissions in terms of RHESSI HXR spectra and flux evolution. From STEREO/EUVI and STEREO/COR1 data the full CME kinematics of the impulsive acceleration phase up to $\sim$4~R$_{\odot}$ is measured with a high time cadence of $\leq$2.5~min. For deriving CME velocity and acceleration we apply and test a new algorithm based on regularization methods. The CME maximum acceleration is achieved at heights $h\leq$~0.4~R$_{\odot}$, the peak velocity at $h\leq$~2.1~R$_{\odot}$ (in one case as small as 0.5~R$_{\odot}$). We find that the CME acceleration profile and the flare energy release as evidenced in the RHESSI hard X-ray flux evolve in a synchronized manner. These results support the ``standard" flare/CME model which is characterized by a feed-back relationship between the large-scale CME acceleration process and the energy release in the associated flare.
\end{abstract}

\section{Introduction}

Solar flares and coronal mass ejections (CMEs) are the most violent phenomena in our solar system. Many aspects of the basic physics of these events are still not well understood. In the ``standard" model it is envisaged that the erupting filament or CME stretches the coronal magnetic field lines to build up a vertical current sheet, where magnetic reconnection sets in to explosively release vast amounts of free magnetic energy, previously stored in the corona in non-potential magnetic fields \citep[for a review see e.g.][]{forbes00}. The released energy goes into plasma heating, acceleration of particles to suprathermal velocities as well as into kinetic energy of the eruption. In this model, a close relation between the kinematics of the CME and the energy release in the associated flare is expected.

A significant fraction of the primary flare energy goes directly into acceleration of fast electrons \citep[e.g.][]{hudson92}. As the accelerated electrons precipitate downward along the newly closed magnetic field lines to the lower lying denser atmospheric layers, they are collisionally stopped and heat and ionize the chromosphere and lower transition region. This can then be observed as enhanced UV and H$\alpha$ radiation. If the beam flux is high enough, they will also emit detectable hard X-rays (HXRs) via nonthermal bremsstrahlung when the electrons scatter off ions of the ambient thermal plasma. Thus, HXR emission provides the most direct indicator of the evolution of the energy release in a flare \citep[e.g.][]{fletcher01}.

Several authors found a correlation between the CME acceleration and the flare soft X-ray emission \citep[][]{zhang01, zhang04,zhang06,vrsnak04,maricic07,vrsnak07-solph}. In a recent case study by \cite{temmer08} an almost synchronized behavior between CME acceleration and flare HXR emission was obtained for two on-disk events. Such results reveal that, for some associated flare-CME events, the reconnection process during the flare is closely related to the CME kinematical evolution. The question remains if it is also possible to relate the flare HXR emission (count rate, spectral parameters) to the CME acceleration magnitude. The Reuven Ramaty High-Energy Solar Spectroscopic Imager \citep[RHESSI;][]{RHESSI} delivers HXR spectra and images at high temporal and spectral resolution, which enables us to study in detail the flare energy release process.

To study the flare energy release in relation to the CME kinematics, the impulsive or main acceleration phase of a CME has to be covered. Since the impulsive acceleration phase of a CME takes place at distances of $R\lesssim$~3~R$_{\odot}$ \citep{macqueen83,stcyr99,vrsnak01,zhang01,temmer08} observations at low coronal heights and of high temporal cadence are required but are only limited available from white light coronagraphs. The Extreme Ultraviolet Imager instruments aboard the twin spacecraft of the Solar Terrestrial Relations Observatory \citep[STEREO;][]{kaiser08} mission have a large field of view and a high time resolution in the 171\AA~passband, perfectly suiting such studies. Together with COR1 observations, the inner coronagraph aboard STEREO, the impulsive acceleration phase of a CME is fully covered from its launch in the low corona up to 4.0~R$_{\odot}$.

In this paper we study and analyze the CME dynamics and HXR emission of the associated flare for three well observed CME/flare events, using the potential of two unprecedented missions, STEREO and RHESSI. The events are selected to be located close to the limb in order to minimize the effect of projection in the CME kinematics. In addition, we apply a new algorithm (based on regularization methods) to derive the CME velocity and acceleration from the measured height-time data. A systematic test of the regularization method as well as a least-squares spline algorithm to a variety of synthetic CME acceleration profiles is performed, in order to evaluate the limitations and uncertainties of both methods.

\section{Data and Methods}

The Sun Earth Connection Coronal and Heliospheric Investigation \citep[SECCHI;][]{howard-stereo08} instrument package is part of each of the STEREO twin spacecraft, STEREO-A(head) and STEREO-B(ehind). It includes among others the Extreme Ultraviolet Imager \citep[EUVI;][]{wuelser04} and the white-light inner coronagraph COR1 \citep[][]{thompson03}. EUVI has a field of view (FoV) of 1.7~R$_{\odot}$ which enables us to follow the erupting CME structure during its initiation and early propagation phase. Combining EUVI with COR1 observations, which have a FoV of 1.4--4.0~R$_{\odot}$, we can derive a complete velocity and acceleration profile for the early CME evolution. The partially overlapping FoVs between EUVI and COR1 allow us to check if the same features are followed in both instruments.

Further advantages are the high time cadence of EUVI and COR1 observations which allow us to study the evolution of impulsive CME events low in the corona. EUVI observes in the EUV 171\AA~passband (Fe\,{\sc x,xi}: $T \sim 1 \times10^6$K) with a nominal cadence of 2.5~min; COR1 has a cadence of 10~min. Each day from 18:00 to 22:00~UT, EUVI and COR1 gather observations with increased cadence, namely 75~sec for EUVI and 5~min for COR1 in order to coordinate with Mauna Loa Solar Observatory as well as during observing campaigns. For the study of the CME kinematics, we combine EUVI~171\AA~filtergrams and total brightness images calculated from COR1 polarization sequence triplets. The level-0 data are properly reduced using SolarSoft routines, all images are background subtracted (for COR1 monthly background is subtracted from each polarization component) and rotated to solar north up.

RHESSI performs imaging spectroscopy of solar flares in the energy range from 3~keV to 17~MeV with high temporal, spatial and spectral resolution \citep{RHESSI}. The flares under study are fully covered by RHESSI observations during their peak phase which enables us to study in detail their HXR spectra and flux evolution in relation to the kinematics of the associated CMEs. RHESSI spectra were integrated over 20~sec during the flare peak with a spectral resolution of 1~keV and fitted with a thermal plus non-thermal power-law component using OSPEX \citep{schwartz02}. From the power-law component we derive the amplitude of the HXR photon spectrum at 50~keV, the spectral index $\gamma$ of the photon spectrum (slope of the power law), and the power in electrons above a cutoff energy of 25~keV for thick-target emission \citep{brown71}. RHESSI images were reconstructed during the flare peak using the CLEAN algorithm \citep{hurford02} including front detectors 2 to 8 (except 5 and 7).

\section{Derivation and test of CME acceleration profiles}\label{method}

For our study the reliable derivation of CME velocity and acceleration profiles from the measured height-time (HT) data is crucial. The central problem is to properly smooth the noisy data and to estimate the impact of measurement errors on the derived quantities, especially on the acceleration profile. This issue received a lot of attention for deriving CME kinematical curves from SOHO/LASCO data \citep[see][]{wen07}. Here, we apply a new type of technique to derive velocity and acceleration profiles from HT measurements, namely an enhanced regularization algorithm originally developed by \cite{kontar04} to invert solar X-ray spectra measured by RHESSI.

The main problem of time derivatives of data with measurement errors is that its finite difference estimate always leads to an amplification of the error. This can be expressed as the sum of discretization (finite time cadence) and propagation errors \citep{groetsch84}. The former error is proportional to the time cadence of the measurements, while the latter is inversely proportional to it. Therefore, the total derivative error always has a minimum \citep{hanke01}. Following \cite{hanke01}, and \cite{kontar05} the regularized method searches for a model-independent velocity and acceleration estimate (regularized solution) with a minimum error, which produces smooth derivatives and avoids additional errors typical of finite differences. In addition, the regularized method of \cite{kontar05} provides the confidence interval for the first and second derivative, i.e.\ for CME velocity and acceleration profiles. The outcome of the regularization algorithm is controlled by two parameters. The number of time bins (which has a smoothing effect) and the regularization ``tweak'', an important parameter that regulates how reliable are the uncertainties of the input data (e.g.\ ``tweak'' of one means that the errors follow a normal distribution without systematics or correlation).

To test the reliability of the results we compare the CME acceleration profiles\footnote{Deriving the acceleration profile, i.e.\ calculating the second derivative, highly intensifies noise on the data. The outcome of the acceleration is critical in order to reliably compare with HXR emission.} derived by the regularization algorithm with those derived from a least-squares spline fit method which is an advanced method already used in previous CME studies \citep{maricic04,vrsnak07-solph}. (Note that we did not compare less sophisticated methods, like direct derivation of the HT data or 3-point Lagrangian interpolation since these would give worse results.) Based on the formulas given in \cite{gallagher03}, we generate synthetic kinematical curves of CME propagation representing different scenarios of CME evolution (impulsive and gradual acceleration profiles).

We then reduce the sampling rate of the generated curve and add some noise. The noise consists of random noise as well as of manually shifting individual data points. With this we aim to take into account the decrease in the dynamic range of CME observations from the inner to the outer corona as well as inconsistencies of tracking features across different wavelength regimes (EUV and white light). From these ``imperfect'' data points the acceleration profile is derived and compared to the acceleration profile from the original ``perfect'' synthetic curve using the regularization and the spline method. To test the stability of the two methods, we simulate HT data with different errors and sampling rates. For both methods the normalized residuals are derived, i.e.\ the differences between the measured HT values and the values resulting from the fitting procedure and the regularized solution, respectively, divided by the error.

In addition to the regularization method we use the spline fit routine which computes least squares splines with equally spaced nodes to the HT data points (SPLFIT in IDL). From the spline fit curve, we numerically derive the second order derivative to determine the acceleration. The smoothing effect of the spline fit is controlled by the number of nodes $n$ \citep[cf.][]{vrsnak07-solph}. To illustrate the way in which this affects the shape of the derived acceleration curve, we present the results of the spline fit for three different numbers of nodes ($n-1,n,n+1$). The nodes are chosen in such a way that the residuals for each of the curves would be equally small.

\begin{figure}
\epsscale{.8}
\plotone{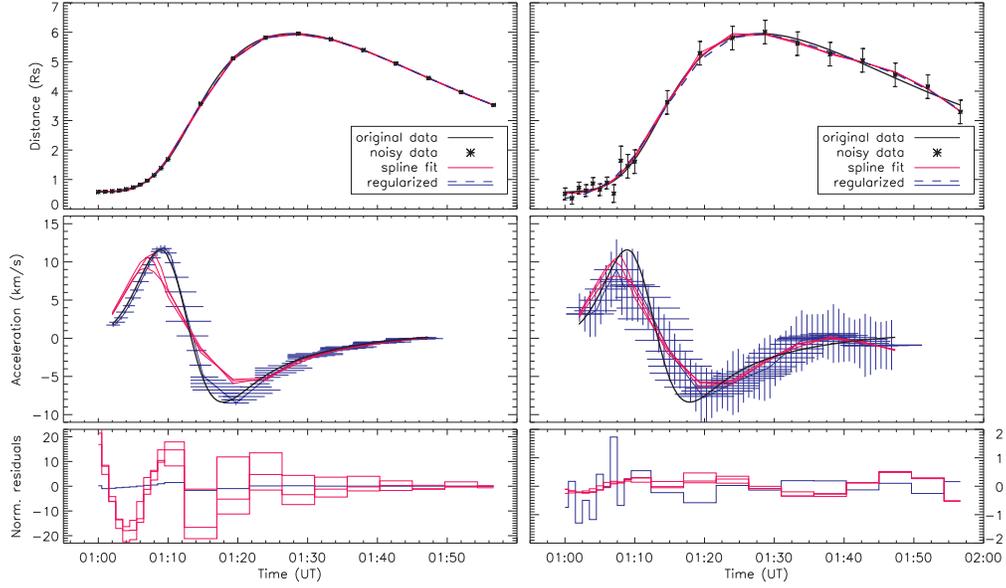}
 \caption{Top: Synthetic curve of CME kinematics (black) together with the data (asterisks), results derived from the regularization tool (dashed blue) and the spline fit (pink). Middle: derived acceleration profile (black: true solution, blue: regularization tool result, pink: spline fit result). Bottom: normalized residuals (we note that in the left panel case the residuals make no sense due to the division by errors of the order of zero). Left: sampling rate is smaller than for the ``true'' curve, but the data are without noise. Right: same sampling rate as left but with noise added to the data. }
 \label{f1-test}
\end{figure}

We first test the reliability of the methods and derive the acceleration from the synthetical HT data without noise (Fig.~\ref{f1-test} left panels). For comparison, the right panels of Fig.~\ref{f1-test} show the results derived from HT data with noise added. We note that the second part of the synthetic curve (decreasing height) is unrealistic for a CME evolution, but the profile was used to provide a challenge to both methods. As can be seen from the left panels, without noise the outcome from the regularization tool represents the acceleration curve better than that from the spline fit with respect to the timing of the acceleration peak and the acceleration duration, though the spline fit method reproduces the HT curve quite well. The peak amplitude of the CME acceleration is well represented by the regularization tool, whereas it is underestimated (by about 20\%) by the spline fit method. In addition, the regularization tool reveals horizontal error bars ($\pm$1.0~min) to account for the uncertainty in timing due to the reduced sampling rate. From the right panels of Fig.~\ref{f1-test} it can be seen that the acceleration curve derived from the spline fit applied on noisy data has only slightly changed, whereas those from the regularization tool did since it considers the information provided by the error bars to yield an adequate error estimation which we do not get from the spline fit method. Strictly speaking, as soon as we add noise to the true HT data, the ``true'' velocity or acceleration cannot be reproduced. So the prime objective of any method is to provide the range of velocities/accelerations where the ``true'' solution should be.

\begin{figure}
\epsscale{.8}
\plotone{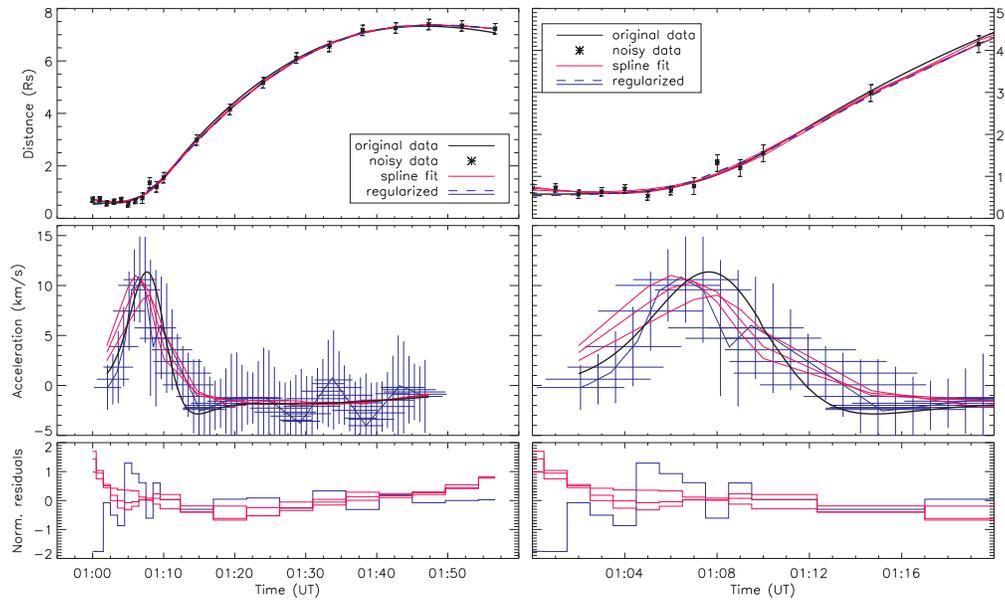}
 \caption{Top to bottom: same as in Fig.~\ref{f1-test}. Left: Impulsively accelerated CME evolution with rather high noise and errors. Right: Close up view for the time range during the peak acceleration phase.}
 \label{f2-test}
\end{figure}

Figure~\ref{f2-test} shows the outcome for a HT curve of an impulsively accelerated CME evolution with large noise added (average error of 0.33~R$_{\odot}$, i.e.\ larger than typically derived from real observations). We see that the acceleration peak is neither well represented by the regularization tool nor by the spline fits. However, the intrinsic errors in time which we get from the regularization tool ($\pm$2.0~min) include the true solution. The overall shape, i.e.\ the duration of acceleration, is well represented by both methods.

\begin{figure}
\epsscale{.8}
\plotone{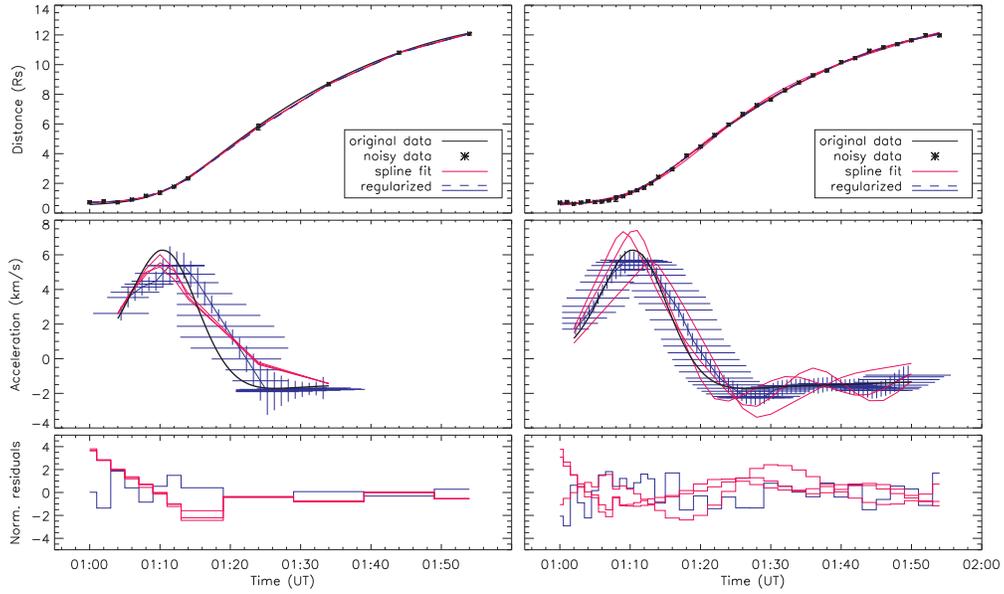}
 \caption{Top to bottom: same as in Fig.~\ref{f1-test}. Realistic scenario of CME evolution applying low noise to the data and error bars comparable to those from the observations. Left: Simulated is a time cadence of data points comparable to the actual observations. Right: same curve but with higher time cadence.}
 \label{f8-test}
\end{figure}

Figure~\ref{f8-test} (left) shows a CME evolution curve with small errors ($\pm$0.07~R$_{\odot}$) and a sampling rate increasing from 2 to 10 minutes, simulating a cadence of measurement points comparable to the actual observations in the present study. The spline fit acceleration varies only little when applying different numbers of nodes (6,7,8) and matches well the peak but not the duration of acceleration. The outcome of the regularization tool for the most part matches the true solution within its error bars (uncertainty in the timing of the acceleration peak about $\pm$3.0 minutes). The right panel of Fig.~\ref{f8-test} shows the same kinematical curve, however, with a higher sampling rate of 1 and 2~min, respectively. The acceleration peak derived from the spline fit using different nodes shows a larger discrepancy. The timing of the acceleration peak calculated from the regularization tool is consistent with the true curve. However, its uncertainty is estimated to about $\pm$4.0~min. To summarize, a sampling rate of $\sim$2~min covering the impulsive acceleration phase reproduces the true curve in a reasonable way. Higher time cadence data produce a better estimate of the acceleration duration. Somewhat surprisingly, the horizontal uncertainties on the acceleration peak are larger when a higher sampling rate is used ($<$2~min). This is due to the fact that small time steps between the HT data points with errors cause larger uncertainties in the subsequent derivative. Thus, the enhancement of the time cadence can only improve the acceleration profiles when also the errors in the CME HT measurements are reduced.

\section{CME acceleration compared to flare energy release}

\begin{figure*}
\epsscale{1.}
\plotone{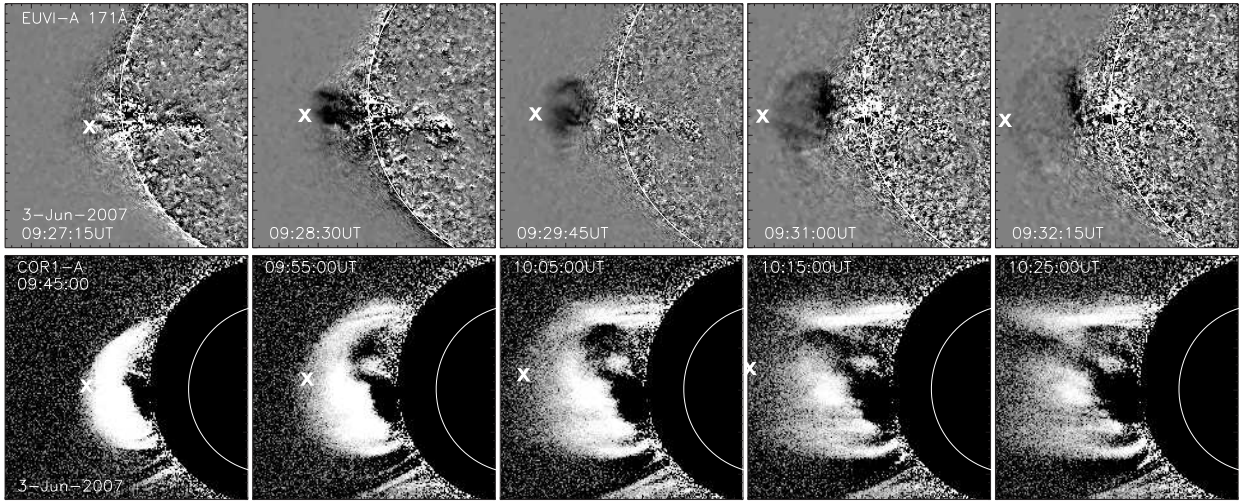}
 \caption{2007 June 3, C5.3 flare/CME event observed with STEREO-A. Sequence of EUVI~171\AA~running difference images and COR1 total brightness images. Crosses indicate the measured leading edge from which the CME kinematics (shown in Fig.~\ref{f1-res}) is derived.}
 \label{f1-obs}
\end{figure*}

\begin{figure*}
\epsscale{1.0}
\plotone{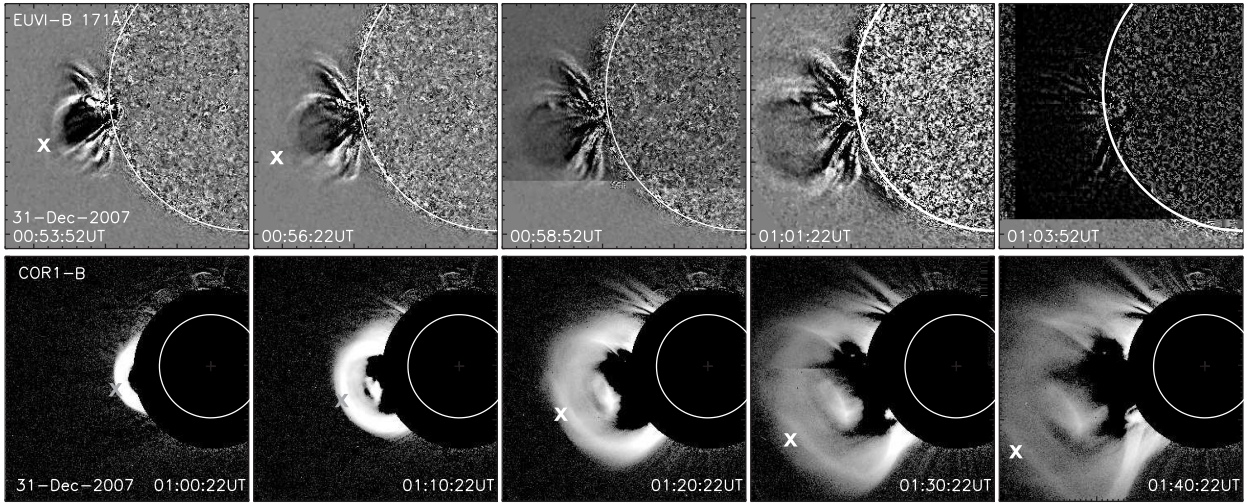}
 \caption{2007 December 31, C8.3 (M2) flare/CME event observed with STEREO-B.
Sequence of EUVI~171\AA~running difference images and COR1 images. Crosses indicate the measured leading edge from which the CME kinematics is derived (Fig.~\ref{f2-res}).}
 \label{f2-obs}
\end{figure*}

\begin{figure*}
\epsscale{1.0}
\plotone{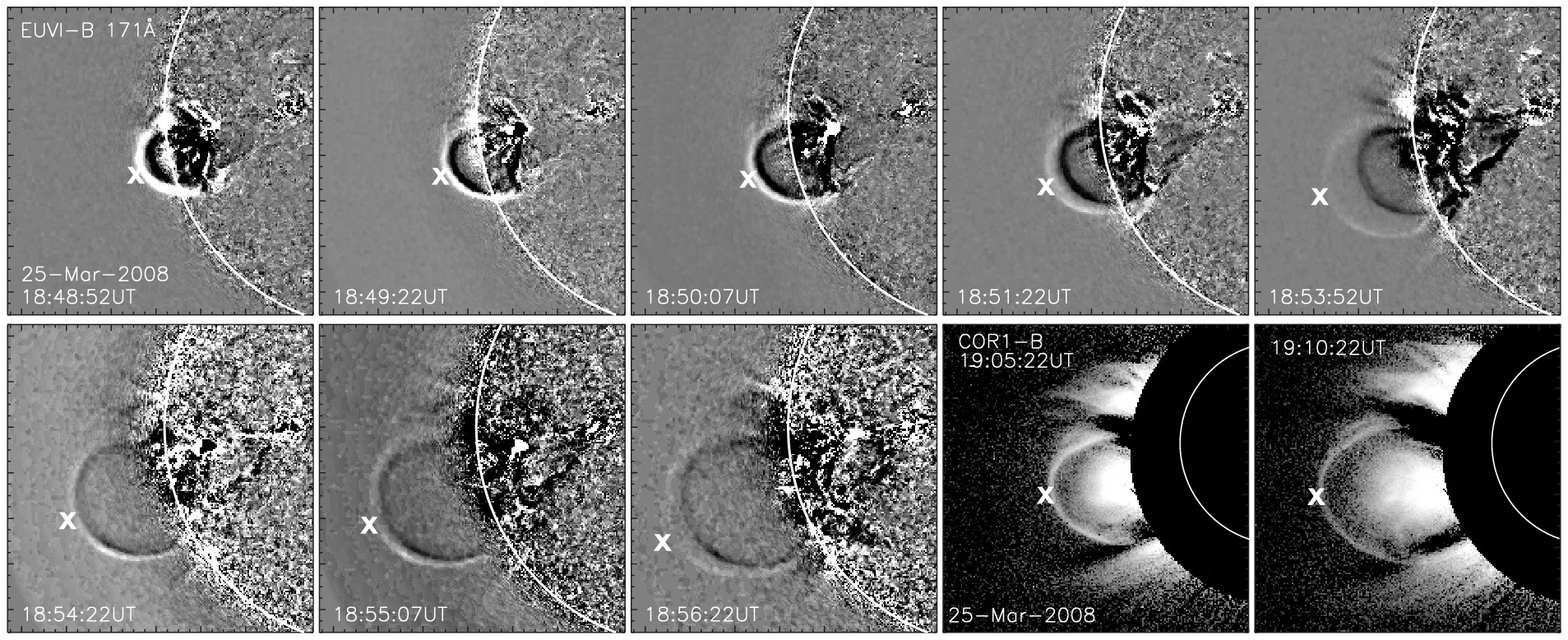}
 \caption{2008 March 25, M1.7 flare/CME event observed with STEREO-B.
Sequence of EUVI~171\AA~running difference images and COR1 images. Crosses indicate the measured leading edge from which the CME kinematics is derived (Fig.~\ref{f3-res}).}
 \label{f3-obs}
\end{figure*}

The flare/CME events under study are: 2007 June 3 (C5.3), 2007 December 31 (C8.3), 2008 March 25 (M1.7). Figures \ref{f1-obs}, \ref{f2-obs}, and \ref{f3-obs} show for the three events the evolution of the erupting CME in the low corona in EUVI~171\AA\, and white light images from COR1. In both instruments we follow the leading edge of the CME as indicated by crosses in the figures. In EUVI~171\AA\, ($T \sim 1 \times10^6$K) typical CME features as the frontal rim and the cavity are observed \citep[see also][]{aschwanden09}. However, the embedded cooler prominence material ($T \sim 1 \times10^4$K), visible in the subsequent coronagraph images, is missing. This is reasonable, since the contrast of prominences against the coronal background drops sharply for lines formed at $T \geq 3 \times10^5$K \citep{noyes72}. The obvious similarity of the morphology and the temporal evolution of the developing/propagating CME structure justifies the link between EUV and coronagraph observations \citep[for a discussion see][]{maricic04}. By combining measurements of the leading edge of the erupting structure from EUVI and COR1 images we derive the kinematics of the CME with typical uncertainties of $\pm$0.02--0.05~R$_{\odot}$ for EUVI and $\pm$0.12~R$_{\odot}$ for COR1. We stress that for each event under study the location of the CME source (estimated by the flare position) is close to the limb ($\pm$10--25$\degree$). Thus, we expect that the influence of projection effects on the derived kinematics is small. We also note that for all events large-scale coronal waves were detected in the EUVI 195\AA~image sequences.

\subsection{2007 June 3: C5.3 flare/CME event}

The 2007 June 3 CME event is associated with a C5.3 GOES class flare at heliographic position S08E67
\citep[EUVI event catalog;][]{aschwanden09}. The position angle of STEREO-A with respect to Earth is 7$\degree$, hence for STEREO-A the CME was observed $\sim$16$\degree$ off the spacecraft plane-of-sky. The event was observed by EUVI with a high cadence of 75~sec.

\begin{figure}
\epsscale{0.4}
\plotone{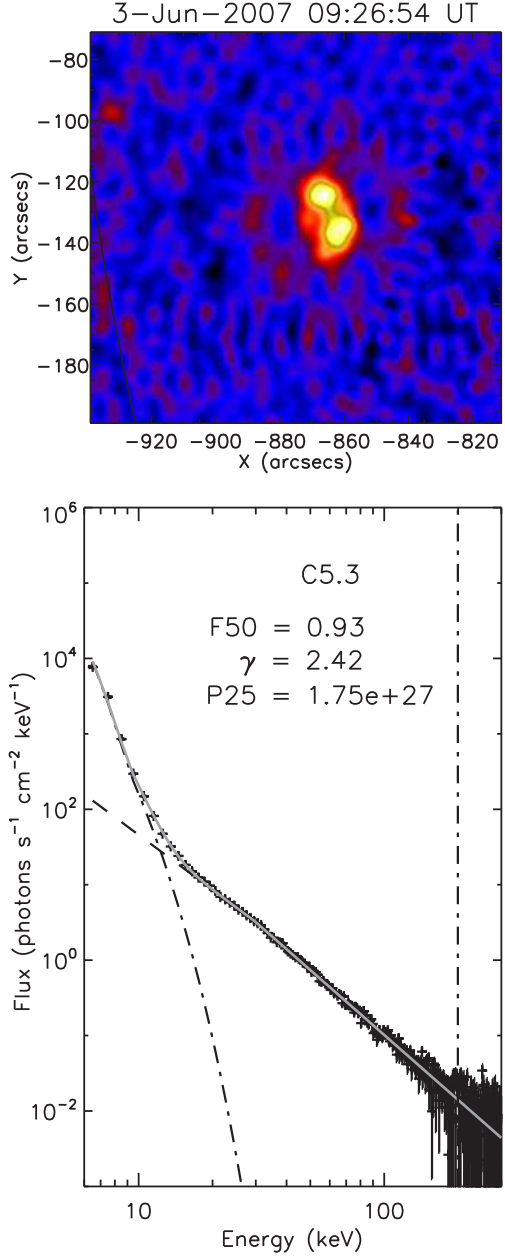}
 \caption{2007 June 3 C5.3 flare. Top: RHESSI 30--50~keV HXR image integrated over 20~sec around the flare peak using front detectors 2 to 8 (except 5 and 7) reconstructed with the CLEAN algorithm. Bottom: RHESSI spectrum and fit components for the energy range 6--200~keV. The isothermal fit is indicated as dashed-dotted line, the non-thermal power-law fit as dashed line, and the sum of both components as thick gray line. The derived parameters, electron flux density at 50~keV $F_{50}$ [photons~s$^{-1}$~cm$^{-2}$~keV$^{-1}$], photon spectral index $\gamma$, and power $P_{25}$ [erg~s$^{-1}$] in electrons above a cutoff energy of 25~keV are given in the legend.}
 \label{new-rh3jun}
\end{figure}

Figure~\ref{new-rh3jun} shows the RHESSI HXR image (top) reconstructed in the energy band 30--50~keV with the CLEAN algorithm \citep{hurford02} together with its spectrum (bottom) integrated over 20~sec around the peak of the HXR emission. The spectrum was fitted with a thermal plus non-thermal model in the energy range 6--200~keV. The HXR source, located close to the eastern limb, is compact with some indication of two footpoints (but not clearly resolved). From the RHESSI spectral fit we derive a photon spectral index $\gamma=2.4$ which indicates a very flat (i.e.\ hard) spectrum and a photon flux density at 50~keV during the event peak, $F_{50}=0.93$ photons~s$^{-1}$~cm$^{-2}$~keV$^{-1}$.

\begin{figure}
\epsscale{0.8}
\plotone{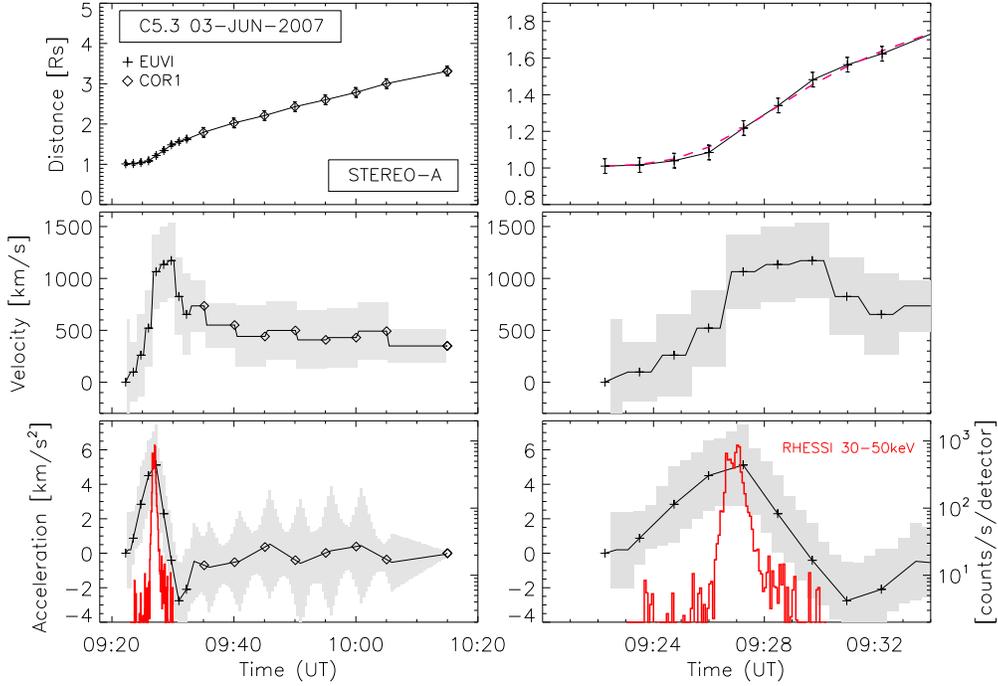}
 \caption{CME kinematics for the 2007 June 3 C5.3 flare/CME event. Top to bottom: CME distance, velocity, and acceleration against time together with the background subtracted flare HXR flux (red). Left: full height range, right: closeup view onto the early evolution phase. The pink dashed line in the right top panel (CME distance-time measurements) shows the regularized solution returned from the inversion technique. The grey shaded area in the CME velocity and acceleration curves indicates the 95\% confidence level.}
 \label{f1-res}
\end{figure}

Figure \ref{f1-res} shows the CME distance, velocity, and acceleration time-profiles for the distance range 1.0--3.3~R$_{\odot}$ (left panels from top to bottom). To enlarge the details on the impulsive acceleration phase, the right panels of Fig.~\ref{f1-res} show the same curves but for the distance range up to 1.7~R$_{\odot}$. For comparison, the RHESSI HXR flux of the associated flare in the energy range 30--50~keV is overplotted in the bottom panels of Fig.~\ref{f1-res}. RHESSI reveals impulsive and powerful HXR emission with a distinct burst of short duration (less than 2 min). The evolution of the acceleration profile of the CME and the evolution of the flare HXR flux are highly synchronized peaking at 09:27~UT with an uncertainty of $\pm$1.25~min (CME) and 09:27:14~UT (RHESSI 30--50~keV HXR flux). The CME reveals a large peak velocity ($\sim$1170~km~s$^{-1}$) and peak acceleration ($\sim$5.1~km~s$^{-2}$), which occur within the EUVI FoV, i.e.\ below 1.7~R$_{\odot}$. The CME accelerates within 8~min up to $\sim$1170~km~s$^{-1}$, then drops within the next 2~min down to a velocity of $\sim$700~km~s$^{-1}$ and further decelerates over the following 10~min until reaching a constant velocity of $\sim$500~km~s$^{-1}$. If one would determine the CME speed solely from coronagraph observations, the velocity peak would be missed.

\subsection{2007 December 31: C8.3 flare/CME event}

The 2007 December 31 CME event is associated with a C8.3 GOES class flare. The source region lies behind the limb at roughly S09E102 \citep{krucker10, dai10} which makes this event special with respect to observations from Earth-view (GOES, RHESSI). From the RHESSI perspective the footpoints of the flare are occulted and we actually observe a coronal HXR source at non-thermal energies, with a good count statistics in the range 20--50 keV. This might be interpreted as Masuda type source \citep{masuda94}. For that day, MESSENGER \citep{solomon01} is at a heliographic longitude of about E165 and observed the non-occulted flare in soft-X rays (similar to GOES). The flux measured by MESSENGER is about 2.8 times higher than the flux observed by GOES which makes the flare actually of M2 class \citep{krucker10}.

\begin{figure}
\epsscale{0.4}
\plotone{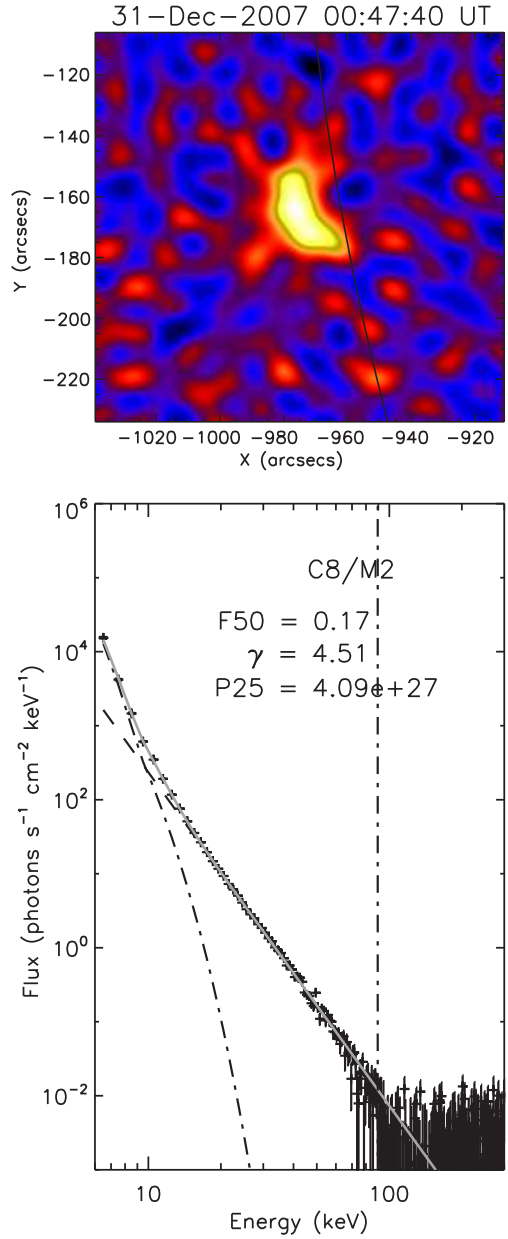}
 \caption{Same as in Fig.~\ref{new-rh3jun} but for the 2007 December 31 C8.3 (M2) flare. The HXR image is integrated over the energy band 20--50~keV.}
 \label{new-rh31dec}
\end{figure}

Figure~\ref{new-rh31dec} shows the RHESSI HXR image reconstructed in the energy band 20--50~keV (top) together with its spectrum (bottom) integrated over 20~sec around the peak of the HXR emission and fitted in the energy range 6--90~keV. The image reveals that the strongest HXR emission comes from above the eastern limb, hence the footpoints of the flare are occulted. This means that the HXR flux entirely originates from the coronal source \citep[e.g.][]{krucker08}. For this event we derive the RHESSI spectral fit parameters as $\gamma=4.5$, which means a softer HXR spectrum compared to the previous event, and $F_{50}=0.17$ photons s$^{-1}$~cm$^{-2}$~keV$^{-1}$. The considerably lower photon flux density and the softer spectrum as compared to the 2007 June 3 event are not surprising since loop-top sources are generally less bright than the footpoints \citep[][]{krucker08b}.

\begin{figure}
\epsscale{0.8}
\plotone{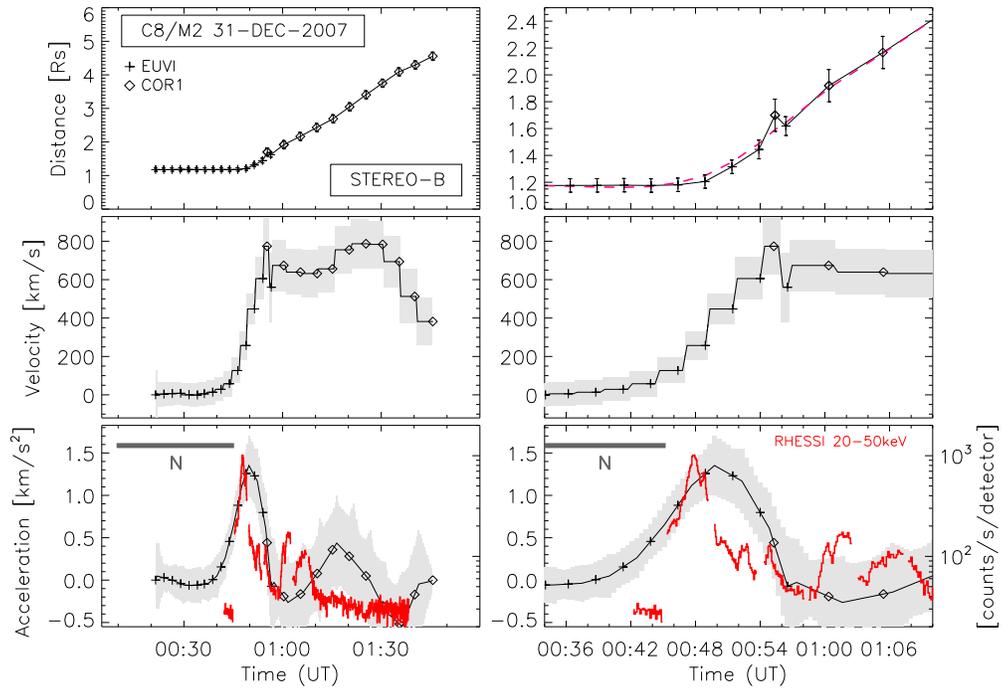}
 \caption{Same as Fig.~\ref{f1-res} but for the 2007 December 31 C8.3 (M2) flare/CME event. The gray bar indicates the time range of RHESSI night (N).}
 \label{f2-res}
\end{figure}

At that day, STEREO-B had a position angle of 23$\degree$ with respect to Earth, hence the CME was observed 11$\degree$ off the STEREO-B plane-of-sky. Figure \ref{f2-res} shows the CME distance-time, velocity, and acceleration profile for the measured distance range 1--5~R$_{\odot}$ (left panels). The closeup view (right panels) shows the distance range up to 2.5~R$_{\odot}$. We stress that for this event one data point (COR1) in the overlapping FoVs of EUVI and COR1 (1.4--1.7~R$_{\odot}$) reveals a shift between the both instruments, but which lies within the applied error bars. The CME reaches a maximum velocity of $\sim$790~km~s$^{-1}$ within $<$10~min, with a maximum acceleration of $\sim$1.3~km~s$^{-2}$. The impulsive acceleration phase of the CME is finished within the EUVI FoV and shows a peak at 00:50~UT ($\pm$2.5~min) which (within the uncertainties) is in accordance with the HXR evolution and peak time at 00:47:48~UT (Fig.~\ref{f2-res}, bottom panels). This is in particular interesting, as in this event we do not observe the HXR emission from the footpoints (which are occulted) but from a coronal source.

\subsection{2008 March 25: M1.7 flare/CME event}

The 2008 March 25 CME event is associated with a M1.7 GOES class flare. The source region lies at S10E87 \citep[EUVI event catalog;][]{aschwanden09} and the position angle of STEREO-B with respect to Earth is 24$\degree$, i.e. the CME propagates about 27$\degree$ off the STEREO-B plane-of-sky. The event is observed with high cadence by EUVI and COR1 (75~sec and 5~min, respectively). RHESSI observed the peak phase of the flare HXR emission, but missed part of the rising phase due to spacecraft night.

\begin{figure}
\epsscale{0.4}
\plotone{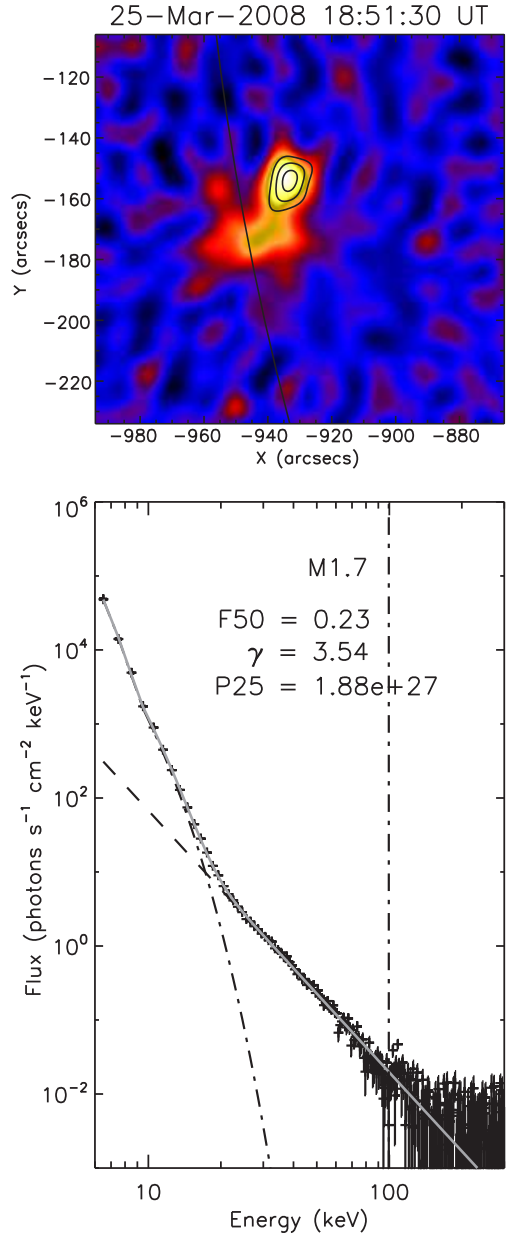}
 \caption{Same as in Fig.~\ref{new-rh3jun} but for the 2008 March 25 M1.7 flare. The HXR image is integrated over the energy band 18--30~keV. Contours show levels of 50\%, 70\%, and 90\% of maximum emission for the energy band 30--50~keV.}
 \label{new-rh25mar}
\end{figure}

Figure~\ref{new-rh25mar} shows the RHESSI HXR image in the energy band 18--30~keV (top) together with its spectrum (bottom) integrated over 20~sec around the peak of the HXR emission. The spectrum was fitted in the energy range 6--100~keV, giving $\gamma=3.5$ and $F_{50}=0.23$ photons s$^{-1}$~cm$^{-2}$~keV$^{-1}$. The image shows a rather compact source of HXR emission revealing a weaker component in the corona in addition to the on-disk footpoint emission, suggesting that the second footpoint lies occulted behind the eastern limb (the energy range 18--30~keV is dominated by non-thermal emission, see spectrum in Fig.~\ref{new-rh25mar}). In EUVI~304\AA~images a bright ejection is observed which may coincide with the coronal RHESSI source.

\begin{figure}
\epsscale{0.8}
\plotone{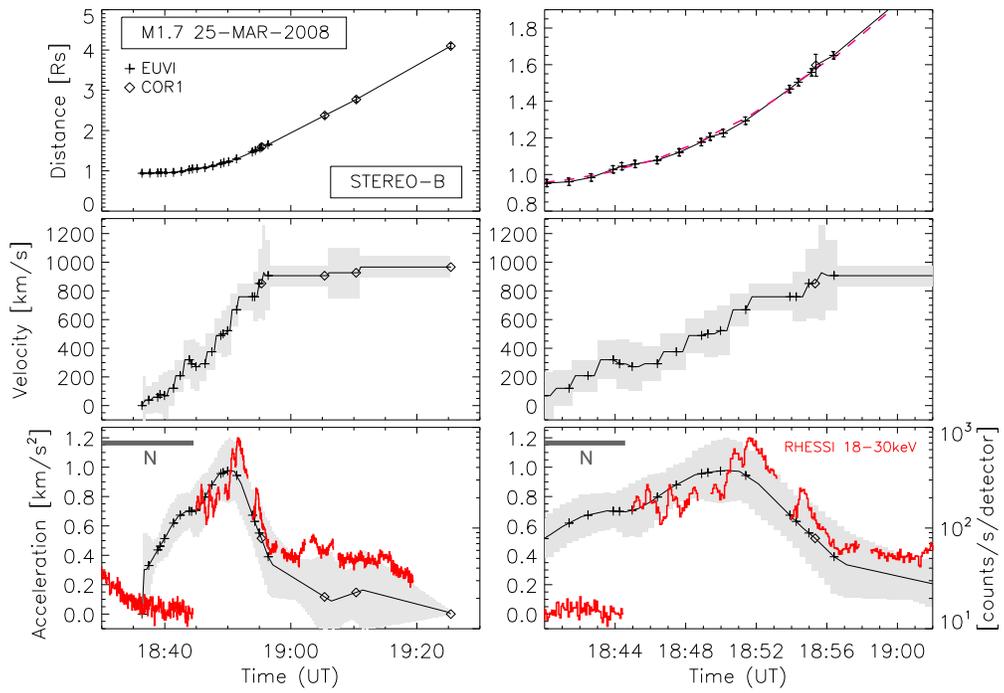}
 \caption{Same as Fig.~\ref{f1-res} but for the 2008 March 25 M1.7 flare/CME event. The gray bar indicates the time range of RHESSI night (N).}
 \label{f3-res}
\end{figure}

Figure \ref{f3-obs} shows the sequence of STEREO-B EUVI running difference and COR1 images. The images reveal a very thin and distinct spherically shaped front of the CME, reminding of the cross-section of a flux-rope torus. This shape is maintained in COR1 white-light images \citep[see also][]{aschwanden09}. The distance-time, velocity, and acceleration profiles are shown in Fig.~\ref{f3-res}, overplotted with the RHESSI HXR flux in the energy range 18--30~keV. The COR1 data point lying in the overlapping FoVs of EUVI and COR1 fits well to the CME HT data measured by EUVI. Due to RHESSI night, the initial flare phase is missed, but the peak is clearly observed. RHESSI comes out of spacecraft night at $\sim$18:44~UT revealing enhanced HXR emission with a maximum at 18:51:34~UT. The CME velocity starts to increase at 18:35~UT, the acceleration reaching a peak at 18:50~UT ($\pm$3.5~min). We note that the acceleration time-profile shows two acceleration steps within the EUVI FoV. The CME reaches a peak velocity of 970~km~s$^{-1}$ within $\sim$20~min with an acceleration maximum of $\sim$1~km~s$^{-2}$.

\section{Summary \& Discussion}

Both the spline fit and the regularization tool determine the acceleration phase of a CME (peak and duration) reasonably well. The main plus of the regularization tool over the spline fit method is the provision of errors in time and amplitude. In all test cases, the true solution was included within this error estimation. We come to the somewhat surprising result that an increased image cadence ($<$2~min) does not reduce the uncertainty in determining the peak time of CME acceleration. This is due to the measurement errors of the CME leading edge. The inconsistencies of the CME leading edge as measured in different wavelength regimes (EUV and white light) as well as the change of intensity over the FoV of a single instrument do not allow us to narrow the measurement errors. The same holds for the acceleration amplitude for which the uncertainties lie in the range between 10\% and 50\%. As a check for the goodness of the obtained velocity and acceleration profiles from the regularization tool, the normalized residuals between the regularized solutions and the distance-time measurements are investigated and found to be small ($\leq$1.5).

Table~\ref{tab1} summarizes the relevant parameters derived from the CME and flare. From the three events under study we obtain that the peak acceleration of the CME is reached within a few minutes after its launch and at low distances of $\leq$0.4~R$_{\odot}$ from the flare site, which we use as an estimate of the CME source region location. We would like to note that all events occurred close to the limb. Thus, projection effects are small, and the derived values should be close to the ``true'' ones. In these events, the impulsive acceleration phase is already finished before the CME is observed in the coronagraph COR1 FoV. The peak velocity of the CME is reached within heights of $<$2.2~R$_{\odot}$ above the source region. The CME acceleration profile and the flare HXR emission are highly synchronized and peak almost simultaneously. The differences between the flare HXR peak and CME acceleration peak are in the range $\Delta t \leq$~2~min. Such differences lie within the limitations for deriving the CME acceleration peak time, which are caused by the finite cadence and measurement errors in the HT measurements. The uncertainties in the peak time of the CME acceleration estimated from the inversion method lie in the range 2--3.5~min (see Section~\ref{method}).

We note that there is no clear relation between CME velocity and GOES class of the associated flare. For the events under study these parameters vary only by a factor of $\sim$2. Statistical studies comparing the flare SXR peak flux and CME peak velocity find a linear correlation coefficient of about $r\sim$0.47 which is not very strong \citep{moon02}. However, in our three events there seems to be a correlation indicated between the CME acceleration peak and the flare HXR peak flux, which both vary over a factor of 5 in the three events under study. Also some relation between the spectral slope of the HXR spectra and the CME acceleration is suggested (harder flare spectrum seems to be related to larger CME acceleration). Of course, three events are inconclusive for such a relationship but our results suggest that this should be tested on a larger event sample.

Especially for the 2007 June 3 event intriguing results are derived. The single HXR burst of only $\sim$2~min duration goes along with a rather high HXR flux density, CME peak acceleration (acceleration phase of $\sim$8~min), and a very hard HXR spectrum. In contrast to that, a relatively low total energy in flare-accelerated electrons and low GOES classification is revealed. Further, the 2007 June 3 CME shows quite unusual behavior with respect to its very strong deceleration profile with values of $-$3~km~s$^{-2}$ and the velocity decreases from 1100~km~s$^{-1}$ to 500~km~s$^{-1}$ very low in the corona ($<$0.5~R$_{\odot}$). Typical CME deceleration values are in the order of $-$0.01 to $-$0.1~km~s$^{-2}$ \citep[e.g.][]{vrs04domi} and are due to the interaction with the solar wind flow. After the magnetic driving force of the eruption ceases, the CME is slowed down by the drag force until its velocity adjusts to the speed of the solar wind \citep[e.g.][and references therein]{cargill04,manoharan06,vrsnak08a}. \cite{reeves06} studied the relationship between the CME acceleration and the thermal energy release rate in a loss-of-equilibrium flux rope model \citep[cf.][]{lin00,lin02}. They found that for high background magnetic fields and fast reconnection rates the evolution of the CME acceleration and flare energy release rate are well correlated. \cite{reeves06} also points out that due to rapid reconnection the formation of a long current sheet in the wake of a CME is prevented and consequently the energy release is inhibited. For the 2007 June 3 event, we therefore assume a localized strong magnetic field from which the CME erupts as well as strong overlying fields. On the one side this yields high reconnection rates over a short time range, on the other side a strong overlying magnetic field would drastically decelerate the CME at low coronal heights.

Well observed flare/CME events close to the limb, such as presented in this study, are less affected by projection effects when analyzing the CME parameters and enable us to reliably determine its kinematics and acceleration values. However, close to the limb the associated flare is observed either as ``classical'' on-disk HXR emitting source (footpoints of flare loops at chromospheric heights) or as coronal loop-top emission only, when the bright footpoint emission is occulted by the solar limb. In partially disk-occulted events, loop-top sources could be studied on a statistical basis \citep[cf.][]{krucker08b} and it is most probable that they are related to the initial location of particle acceleration \citep[e.g.][]{krucker08}. \cite{battaglia06} studied five RHESSI flares where HXR emission from both footpoints and the loop-top was observed. They found that the coronal and footpoint sources are well correlated with respect to their temporal and spectral evolution. Our results show in fact that the CME acceleration peak phase is well correlated with both types of HXR sources, i.e.\ coronal and footpoint HXR emission of the associated flare.

In the present study we found for all three events under study, that the impulsive acceleration of the CME is finished at distances of $\leq$0.4~R$_{\odot}$ above the solar surface, i.e.\ within the EUVI FoV and before reaching the COR1 FoV. Hence, from coronagraphic observations only, the early CME evolution may be considerably misinterpreted. \cite{vrsnak90} and \cite{chen03} obtained from analytical models, in which CMEs are treated as toroidal field structures, that the peak of the acceleration should be reached when the half-separation of the CME footpoints is comparable to the CME height above the flare site. For the 2008 March 25 event the CME footpoint separation can be derived with adequate accuracy (see Fig.~\ref{f2-obs}). We obtained a CME half-separation of $\sim$210$\pm$30~Mm, which is comparable to its peak height of $\sim$280~Mm.

Our results point to a possible relation between the flare HXR spectra and the CME acceleration (the event with the highest CME acceleration has the highest HXR flux density and flattest spectrum) whereas the relation to the peak power in electrons and the GOES soft X-ray peak flux seems weak. \cite{maricic07} pointed out that the reconnection rate is more relevant for the CME acceleration than a strong heating and non-thermal particle acceleration. From our results it seems that the efficiency of accelerating particles to high energies is better correlated to the CME acceleration than the total number and energy in electrons (though we note that more statistics on this is needed). Such results may provide new constraints on electron acceleration models and its magnetic geometry as the particle acceleration mechanism has to go along with a rapid closing of magnetic field lines which is needed to drive the CME.

\section{Conclusions}

\begin{figure}
\epsscale{0.4}
\plotone{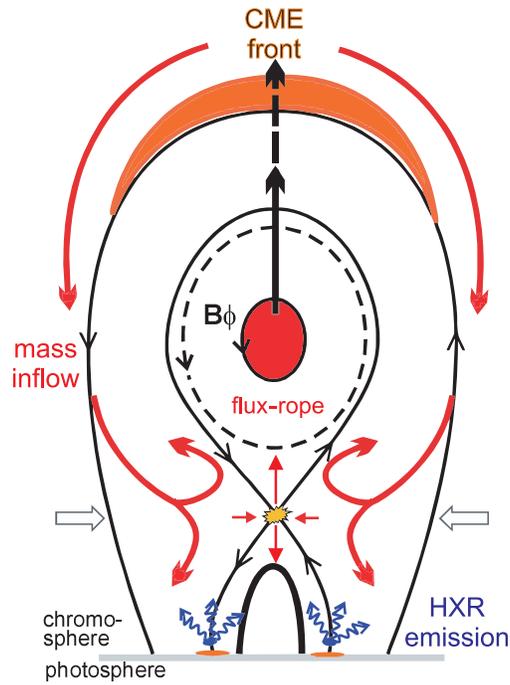}
 \caption{Cartoon illustrating the flare-CME feedback mechanism between the large-scale CME dynamics and small-scale flare processes. The upward moving CME evacuates the area in its wake, boosting mass inflow from aside into the reconnection region. The more mass and frozen-in magnetic field is transported into the region, the higher is the magnetic reconnection rate leading to larger flare energy release and to more efficient acceleration of particles. The successive closing of magnetic field lines due to reconnection increases the poloidal flux B$_{\phi}$ in the eruption, which leads to a stronger upward oriented magnetic driving force (Lorentz force).}
 \label{f5}
\end{figure}

The present study based on high cadence data from STEREO-EUVI, STEREO-COR1, and RHESSI evidences a very close relation between flares and CMEs. From three well observed impulsive events that occurred close to the limb, we derive that the CME acceleration profile and energy release of the associated flare evolve in a synchronized manner. This supports the ``standard" flare/CME model which predicts a feed-back relationship between the large-scale CME acceleration and the energy release process in the associated flare \cite[e.g.][]{lin04a,maricic07,temmer08,vrsnak08b}: After the magnetic structure looses equilibrium and starts rising, a current sheet is formed below the rising structure (presumably a flux-rope), becoming a site of magnetic field reconnection. The reconnection has two important consequences for
the CME acceleration: it reduces the downward-acting tension of the overlying field \citep{lin04a}, and it supplies additional poloidal flux to the flux rope \citep{vrs08angeo}, which is considered to be the main driver of the eruption \citep[e.g.][]{chen89,vrsnak90,chen96,kliem06,subramanian07}. On the other hand, the upward moving CME drives mass inflow into the current sheet, driving further magnetic reconnection and particle acceleration (see the cartoon in Fig.~\ref{f5}). Thus, reconnection directly relates the CME acceleration and the energy release in the associated flare, leading to the close synchronization of these two phenomena.

The results from the presented study provide strong evidence for the feed-back mechanism between the flare energy release and the CME acceleration. In our study, all events were fast eruptions, hence, we can not draw conclusions on other types of eruptions (failed eruptions or gradual CMEs without flares). According to model calculations by \cite{reeves06} a good correlation (within 2~min) between the flux rope acceleration and thermal energy release rate is expected for fast reconnection events with high background magnetic fields. This should relate to impulsive and strong events, whereas no synchronization is expected for weak and gradual events. Finally, we note that high cadence observations of CMEs in non-coronagraphic images enable us to study in detail the CME impulsive acceleration phase revealing peak acceleration as high as $\sim$5~km~s$^{-2}$ even for C-class flare events. We also note that the height of the CME peak acceleration in such impulsive events is much lower ($\leq$0.4~R$_{\odot}$) than previously assumed.

\acknowledgments The authors appreciate and thank the anonymous referee for his/her constructive comments and suggestions. We thank the STEREO and RHESSI teams for their open data policy. M.T. is a recipient of an APART-fellowship of the Austrian Academy of Sciences at the Institute of Physics, University of Graz (APART 11262). A.V. acknowledges the Austrian Science Fund (FWF): [P20867-N16]. E.P.K. work was supported by a STFC Rolling Grant and STFC Advanced Fellowship. The European Community's Seventh Framework Programme (FP7/2007-2013) under grant agreement no.218816 (SOTERIA) is acknowledged.

\bibliographystyle{apj}

\begin{thebibliography}{54}
\expandafter\ifx\csname natexlab\endcsname\relax\def\natexlab#1{#1}\fi

\bibitem[{{Aschwanden} {et~al.}(2009){Aschwanden}, {Wuelser}, {Nitta}, \&
  {Lemen}}]{aschwanden09}
{Aschwanden}, M.~J., {Wuelser}, J.~P., {Nitta}, N.~V., \& {Lemen}, J.~R. 2009,
  \solphys, 256, 3

\bibitem[{{Battaglia} \& {Benz}(2006)}]{battaglia06}
{Battaglia}, M., \& {Benz}, A.~O. 2006, \aap, 456, 751

\bibitem[{{Brown}(1971)}]{brown71}
{Brown}, J.~C. 1971, \solphys, 18, 489

\bibitem[{{Cargill}(2004)}]{cargill04}
{Cargill}, P.~J. 2004, \solphys, 221, 135

\bibitem[{{Chen}(1989)}]{chen89}
{Chen}, J. 1989, \apj, 338, 453

\bibitem[{{Chen}(1996)}]{chen96}
---. 1996, \jgr, 101, 27499

\bibitem[{{Chen} \& {Krall}(2003)}]{chen03}
{Chen}, J., \& {Krall}, J. 2003, Journal of Geophysical Research (Space
  Physics), 108, 2

\bibitem[{{Dai} {et~al.}(2010){Dai}, {Auch{\`e}re}, {Vial}, {Tang}, \&
  {Zong}}]{dai10}
{Dai}, Y., {Auch{\`e}re}, F., {Vial}, J., {Tang}, Y.~H., \& {Zong}, W.~G. 2010,
  \apj, 708, 913

\bibitem[{{Fletcher} \& {Hudson}(2001)}]{fletcher01}
{Fletcher}, L., \& {Hudson}, H. 2001, \solphys, 204, 69

\bibitem[{{Forbes}(2000)}]{forbes00}
{Forbes}, T.~G. 2000, \jgr, 105, 23153

\bibitem[{{Gallagher} {et~al.}(2003){Gallagher}, {Lawrence}, \&
  {Dennis}}]{gallagher03}
{Gallagher}, P.~T., {Lawrence}, G.~R., \& {Dennis}, B.~R. 2003, \apjl, 588, L53

\bibitem[{{Groetsch}(1984)}]{groetsch84}
{Groetsch}, C.~W. 1984, Research notes in mathematics, 105

\bibitem[{{Hanke} \& {Scherzer}(2001)}]{hanke01}
{Hanke}, M., \& {Scherzer}, O. 2001, Am. Math. Mon., 108, 512

\bibitem[{{Howard} {et~al.}(2008){Howard}, {Moses}, {Vourlidas}, {Newmark},
  {Socker}, {Plunkett}, {Korendyke}, {Cook}, {Hurley}, {Davila}, {Thompson},
  {St Cyr}, {Mentzell}, {Mehalick}, {Lemen}, {Wuelser}, {Duncan}, {Tarbell},
  {Wolfson}, {Moore}, {Harrison}, {Waltham}, {Lang}, {Davis}, {Eyles},
  {Mapson-Menard}, {Simnett}, {Halain}, {Defise}, {Mazy}, {Rochus}, {Mercier},
  {Ravet}, {Delmotte}, {Auchere}, {Delaboudiniere}, {Bothmer}, {Deutsch},
  {Wang}, {Rich}, {Cooper}, {Stephens}, {Maahs}, {Baugh}, {McMullin}, \&
  {Carter}}]{howard-stereo08}
{Howard}, R.~A., {et~al.} 2008, Space Science Reviews, 136, 67

\bibitem[{{Hudson} {et~al.}(1992){Hudson}, {Acton}, {Hirayama}, \&
  {Uchida}}]{hudson92}
{Hudson}, H.~S., {Acton}, L.~W., {Hirayama}, T., \& {Uchida}, Y. 1992, \pasj,
  44, L77

\bibitem[{{Hurford} {et~al.}(2002){Hurford}, {Schmahl}, {Schwartz}, {Conway},
  {Aschwanden}, {Csillaghy}, {Dennis}, {Johns-Krull}, {Krucker}, {Lin},
  {McTiernan}, {Metcalf}, {Sato}, \& {Smith}}]{hurford02}
{Hurford}, G.~J., {et~al.} 2002, \solphys, 210, 61

\bibitem[{{Kaiser} {et~al.}(2008){Kaiser}, {Kucera}, {Davila}, {St.~Cyr},
  {Guhathakurta}, \& {Christian}}]{kaiser08}
{Kaiser}, M.~L., {Kucera}, T.~A., {Davila}, J.~M., {St.~Cyr}, O.~C.,
  {Guhathakurta}, M., \& {Christian}, E. 2008, Space Science Reviews, 136, 5

\bibitem[{{Kliem} \& {T{\"o}r{\"o}k}(2006)}]{kliem06}
{Kliem}, B., \& {T{\"o}r{\"o}k}, T. 2006, Physical Review Letters, 96, 255002

\bibitem[{{Kontar} \& {MacKinnon}(2005)}]{kontar05}
{Kontar}, E.~P., \& {MacKinnon}, A.~L. 2005, \solphys, 227, 299

\bibitem[{{Kontar} {et~al.}(2004){Kontar}, {Piana}, {Massone}, {Emslie}, \&
  {Brown}}]{kontar04}
{Kontar}, E.~P., {Piana}, M., {Massone}, A.~M., {Emslie}, A.~G., \& {Brown},
  J.~C. 2004, \solphys, 225, 293

\bibitem[{{Krucker} {et~al.}(2008){Krucker}, {Battaglia}, {Cargill},
  {Fletcher}, {Hudson}, {MacKinnon}, {Masuda}, {Sui}, {Tomczak}, {Veronig},
  {Vlahos}, \& {White}}]{krucker08}
{Krucker}, S., {et~al.} 2008, \aapr, 16, 155

\bibitem[{{Krucker} {et~al.}(2010){Krucker}, {Hudson}, {Glesener}, {White},
  {Masuda}, {Wuelser}, \& {Lin}}]{krucker10}
{Krucker}, S., {Hudson}, H.~S., {Glesener}, L., {White}, S.~M., {Masuda}, S.,
  {Wuelser}, J.-P., \& {Lin}, R.~P. 2010, \apj, submitted

\bibitem[{{Krucker} \& {Lin}(2008)}]{krucker08b}
{Krucker}, S., \& {Lin}, R.~P. 2008, \apj, 673, 1181

\bibitem[{{Lin}(2002)}]{lin02}
{Lin}, J. 2002, Chinese Journal of Astronomy and Astrophysics, 2, 539

\bibitem[{{Lin}(2004)}]{lin04a}
---. 2004, \solphys, 219, 169

\bibitem[{{Lin} \& {Forbes}(2000)}]{lin00}
{Lin}, J., \& {Forbes}, T.~G. 2000, \jgr, 105, 2375

\bibitem[{{Lin} {et~al.}(2002){Lin}, {Dennis}, {Hurford}, {Smith}, \& {the
  RHESSI team}}]{RHESSI}
{Lin}, R.~P., {Dennis}, B.~R., {Hurford}, G.~J., {Smith}, D.~M., \& {the RHESSI
  team}. 2002, \solphys, 210, 3

\bibitem[{{MacQueen} \& {Fisher}(1983)}]{macqueen83}
{MacQueen}, R.~M., \& {Fisher}, R.~R. 1983, \solphys, 89, 89

\bibitem[{{Manoharan}(2006)}]{manoharan06}
{Manoharan}, P.~K. 2006, \solphys, 235, 345

\bibitem[{{Mari{\v c}i{\'c}} {et~al.}(2004){Mari{\v c}i{\'c}}, {Vr{\v s}nak},
  {Stanger}, \& {Veronig}}]{maricic04}
{Mari{\v c}i{\'c}}, D., {Vr{\v s}nak}, B., {Stanger}, A.~L., \& {Veronig}, A.
  2004, \solphys, 225, 337

\bibitem[{{Mari{\v c}i{\'c}} {et~al.}(2007){Mari{\v c}i{\'c}}, {Vr{\v s}nak},
  {Stanger}, {Veronig}, {Temmer}, \& {Ro{\v s}a}}]{maricic07}
{Mari{\v c}i{\'c}}, D., {Vr{\v s}nak}, B., {Stanger}, A.~L., {Veronig}, A.~M.,
  {Temmer}, M., \& {Ro{\v s}a}, D. 2007, \solphys, 241, 99

\bibitem[{{Masuda} {et~al.}(1994){Masuda}, {Kosugi}, {Hara}, {Tsuneta}, \&
  {Ogawara}}]{masuda94}
{Masuda}, S., {Kosugi}, T., {Hara}, H., {Tsuneta}, S., \& {Ogawara}, Y. 1994,
  \nat, 371, 495

\bibitem[{{Moon} {et~al.}(2002){Moon}, {Choe}, {Wang}, {Park}, {Gopalswamy},
  {Yang}, \& {Yashiro}}]{moon02}
{Moon}, Y.-J., {Choe}, G.~S., {Wang}, H., {Park}, Y.~D., {Gopalswamy}, N.,
  {Yang}, G., \& {Yashiro}, S. 2002, \apj, 581, 694

\bibitem[{{Noyes} {et~al.}(1972){Noyes}, {Dupree}, {Huber}, {Parkinson},
  {Reeves}, \& {Withbroe}}]{noyes72}
{Noyes}, R.~W., {Dupree}, A.~K., {Huber}, M.~C.~E., {Parkinson}, W.~H.,
  {Reeves}, E.~M., \& {Withbroe}, G.~L. 1972, \apj, 178, 515

\bibitem[{{Reeves}(2006)}]{reeves06}
{Reeves}, K.~K. 2006, \apj, 644, 592

\bibitem[{{Schwartz} {et~al.}(2002){Schwartz}, {Csillaghy}, {Tolbert},
  {Hurford}, {Mc Tiernan}, \& {Zarro}}]{schwartz02}
{Schwartz}, R.~A., {Csillaghy}, A., {Tolbert}, A.~K., {Hurford}, G.~J., {Mc
  Tiernan}, J., \& {Zarro}, D. 2002, \solphys, 210, 165

\bibitem[{{Solomon} {et~al.}(2001){Solomon}, {McNutt}, {Gold}, {Acu{\~n}a},
  {Baker}, {Boynton}, {Chapman}, {Cheng}, {Gloeckler}, {Head}, {Krimigis},
  {McClintock}, {Murchie}, {Peale}, {Phillips}, {Robinson}, {Slavin}, {Smith},
  {Strom}, {Trombka}, \& {Zuber}}]{solomon01}
{Solomon}, S.~C., {et~al.} 2001, \planss, 49, 1445

\bibitem[{{St.~Cyr} {et~al.}(1999){St.~Cyr}, {Burkepile}, {Hundhausen}, \&
  {Lecinski}}]{stcyr99}
{St.~Cyr}, O.~C., {Burkepile}, J.~T., {Hundhausen}, A.~J., \& {Lecinski}, A.~R.
  1999, \jgr, 104, 12493

\bibitem[{{Subramanian} \& {Vourlidas}(2007)}]{subramanian07}
{Subramanian}, P., \& {Vourlidas}, A. 2007, \aap, 467, 685

\bibitem[{{Temmer} {et~al.}(2008){Temmer}, {Veronig}, {Vr{\v s}nak},
  {Ryb{\'a}k}, {G{\"o}m{\"o}ry}, {Stoiser}, \& {Mari{\v c}i{\'c}}}]{temmer08}
{Temmer}, M., {Veronig}, A.~M., {Vr{\v s}nak}, B., {Ryb{\'a}k}, J.,
  {G{\"o}m{\"o}ry}, P., {Stoiser}, S., \& {Mari{\v c}i{\'c}}, D. 2008, \apjl,
  673, L95

\bibitem[{{Thompson} {et~al.}(2003){Thompson}, {Davila}, {Fisher}, {Orwig},
  {Mentzell}, {Hetherington}, {Derro}, {Federline}, {Clark}, {Chen},
  {Tveekrem}, {Martino}, {Novello}, {Wesenberg}, {StCyr}, {Reginald}, {Howard},
  {Mehalick}, {Hersh}, {Newman}, {Thomas}, {Card}, \& {Elmore}}]{thompson03}
{Thompson}, W.~T., {et~al.} 2003, in Presented at the Society of Photo-Optical
  Instrumentation Engineers (SPIE) Conference, Vol. 4853, Society of
  Photo-Optical Instrumentation Engineers (SPIE) Conference Series, ed. S.~L.
  {Keil} \& S.~V. {Avakyan}, 1--11

\bibitem[{{Vr{\v s}nak}(1990)}]{vrsnak90}
{Vr{\v s}nak}, B. 1990, \solphys, 129, 295

\bibitem[{{Vr{\v s}nak}(2001)}]{vrsnak01}
---. 2001, \jgr, 106, 25249

\bibitem[{{Vr{\v s}nak}(2008)}]{vrs08angeo}
---. 2008, Annales Geophysicae, 26, 3089

\bibitem[{{Vr{\v s}nak} \& {Cliver}(2008)}]{vrsnak08b}
{Vr{\v s}nak}, B., \& {Cliver}, E.~W. 2008, \solphys, 253, 215

\bibitem[{{Vr{\v s}nak} {et~al.}(2007){Vr{\v s}nak}, {Mari{\v c}i{\'c}},
  {Stanger}, {Veronig}, {Temmer}, \& {Ro{\v s}a}}]{vrsnak07-solph}
{Vr{\v s}nak}, B., {Mari{\v c}i{\'c}}, D., {Stanger}, A.~L., {Veronig}, A.~M.,
  {Temmer}, M., \& {Ro{\v s}a}, D. 2007, \solphys, 241, 85

\bibitem[{{Vr{\v s}nak} {et~al.}(2004){Vr{\v s}nak}, {Ru{\v z}djak}, {Sudar},
  \& {Gopalswamy}}]{vrsnak04}
{Vr{\v s}nak}, B., {Ru{\v z}djak}, D., {Sudar}, D., \& {Gopalswamy}, N. 2004,
  \aap, 423, 717

\bibitem[{{Vr{\v s}nak} {et~al.}(2004a){Vr{\v s}nak}, {Ru{\v z}djak}, {Sudar},
  \& {Gopalswamy}}]{vrs04domi}
---. 2004a, \aap, 423, 717

\bibitem[{{Vr{\v s}nak} {et~al.}(2008){Vr{\v s}nak}, {Vrbanec}, \& {{\v
  C}alogovi{\'c}}}]{vrsnak08a}
{Vr{\v s}nak}, B., {Vrbanec}, D., \& {{\v C}alogovi{\'c}}, J. 2008, \aap, 490,
  811

\bibitem[{{Wen} {et~al.}(2007){Wen}, {Maia}, \& {Wang}}]{wen07}
{Wen}, Y., {Maia}, D.~J.~F., \& {Wang}, J. 2007, \apj, 657, 1117

\bibitem[{{Wuelser} {et~al.}(2004){Wuelser}, {Lemen}, {Tarbell}, {Wolfson},
  {Cannon}, {Carpenter}, {Duncan}, {Gradwohl}, {Meyer}, {Moore}, {Navarro},
  {Pearson}, {Rossi}, {Springer}, {Howard}, {Moses}, {Newmark},
  {Delaboudiniere}, {Artzner}, {Auchere}, {Bougnet}, {Bouyries}, {Bridou},
  {Clotaire}, {Colas}, {Delmotte}, {Jerome}, {Lamare}, {Mercier}, {Mullot},
  {Ravet}, {Song}, {Bothmer}, \& {Deutsch}}]{wuelser04}
{Wuelser}, J.-P., {et~al.} 2004, in Presented at the Society of Photo-Optical
  Instrumentation Engineers (SPIE) Conference, Vol. 5171, Society of
  Photo-Optical Instrumentation Engineers (SPIE) Conference Series, ed.
  S.~{Fineschi} \& M.~A. {Gummin}, 111--122

\bibitem[{{Zhang} \& {Dere}(2006)}]{zhang06}
{Zhang}, J., \& {Dere}, K.~P. 2006, \apj, 649, 1100

\bibitem[{{Zhang} {et~al.}(2001){Zhang}, {Dere}, {Howard}, {Kundu}, \&
  {White}}]{zhang01}
{Zhang}, J., {Dere}, K.~P., {Howard}, R.~A., {Kundu}, M.~R., \& {White}, S.~M.
  2001, \apj, 559, 452

\bibitem[{{Zhang} {et~al.}(2004){Zhang}, {Dere}, {Howard}, \&
  {Vourlidas}}]{zhang04}
{Zhang}, J., {Dere}, K.~P., {Howard}, R.~A., \& {Vourlidas}, A. 2004, \apj,
  604, 420

\end{thebibliography}

\newpage

\newpage

\begin{table}[ht]
\caption{Summary of the CME and flare characteristics for each event. We give the date, the GOES flare class, the difference $\Delta$t~[min] between CME peak acceleration and flare HXR peak, the derived CME peak velocity v$_{\rm max}$ [km~s$^{-1}$] and peak acceleration a$_{\rm max}$ [km~s$^{-2}$] with errors (95\% confidence level), the CME height $h$ [R$_{\odot}$] from the source region, the electron flux density at 50~keV, $F_{50}$ [photons~s$^{-1}$~cm$^{-2}$~keV$^{-1}$], the flare HXR spectral slope $\gamma$, and the total power in electrons $P_{25}$ [10$^{27}$~erg~s$^{-1}$]. \label{tab1}}
\bigskip
 \begin{tabular}{llllllll}
\hline
Date & Class & $\Delta t$ & $v_{\rm max}$ ($h$)  & $a_{\rm max}$ ($h$)   & $F_{50}$ & $\gamma$ &  $P_{25}$ \\
\hline\hline
\medskip
03-Jun-07 & C5.3 &  0.1 & 1171$\pm$359 (0.48) & 5.1$\pm$2.4 (0.26)       & 0.93      & 2.4     & 1.75  \\
\medskip
31-Dec-07 & C8/M2 & 2.0 & 785$\pm$125 (2.10) & 1.3$\pm$0.4 (0.25)        & 0.17      & 4.5     & 4.09 \\
\medskip
25-Mar-08 & M1.7 &  1.5 & 967$\pm$173 (1.97) & 1.0$\pm$0.2 (0.40)        & 0.23      & 3.5     & 1.88  \\
\hline
\end{tabular}
 \end{table}

\end{document}